\pdfoutput=1
\documentclass[sigconf]{acmart}

\usepackage{booktabs}
\usepackage{graphicx}
\usepackage{subcaption}
\usepackage{adjustbox}
\usepackage{todonotes}
\usepackage{algorithm}
\usepackage{algorithmic}
\usepackage{placeins}
\usepackage{balance}

\setcopyright{none}

\renewcommand\footnotetextcopyrightpermission[1]{}
\begin{document}
\title{All You Need is Ratings: A Clustering Approach to\\ Synthetic Rating Datasets Generation}

\author{Diego Monti}
\orcid{0000-0002-3821-5379}
\affiliation{Politecnico di Torino}
\email{diego.monti@polito.it}

\author{Giuseppe Rizzo}
\orcid{0000-0003-0083-813X}
\affiliation{LINKS Foundation}
\email{giuseppe.rizzo@linksfoundation.com}

\author{Maurizio Morisio}
\orcid{0000-0001-7362-906X}
\affiliation{Politecnico di Torino}
\email{maurizio.morisio@polito.it}

\begin{abstract}
The public availability of collections containing user preferences is of vital importance for performing offline evaluations in the field of recommender systems. However, the number of rating datasets is limited because of the costs required for their creation and the fear of violating the privacy of the users by sharing them. For this reason, numerous research attempts investigated the creation of synthetic collections of ratings using generative approaches. Nevertheless, these datasets are usually not reliable enough for conducting an evaluation campaign. In this paper, we propose a method for creating synthetic datasets with a configurable number of users that mimic the characteristics of already existing ones. We empirically validated the proposed approach by exploiting the synthetic datasets for evaluating different recommenders and by comparing the results with the ones obtained using real datasets.
\end{abstract}


\maketitle

\section{Introduction}
\label{sec:introduction}

It is widely known that novel recommendation approaches should be evaluated in the context of online experiments involving human subjects in order to obtain reasonably robust results about their performance~\cite{Herlocker2004}. Nevertheless, most of the studies available in literature support their conclusions with offline trails relying on the preferences of users collected without considering the algorithms under investigation~\cite{Gunawardana2015}. Despite the possible weaknesses of this approach~\cite{Said2014}, offline experiments are extremely popular among researchers because of their limited costs and the theoretical reproducibility of their results. In industry, they are usually considered a powerful tool for pruning the number of possible recommender systems that need to be tested with real users, thus mitigating the economical impact of eventual failures.

It is necessary to rely on a collection of user preferences obtained in a particular domain to perform an offline experiment. For example, the MovieLens datasets represent a popular choice for conducting an offline evaluation in the field of movie recommender systems~\cite{Harper2015}. Nevertheless, the number and the variety of publicly available rating datasets is often limited, especially in less mainstream domains~\cite{Tso2006}. It is possible to identify different causes for this problem. For example, the companies capable of collecting rating datasets are usually reluctant to share them, because of the fear of violating the privacy of their users or of exposing commercially sensible data to their competitors. On the other hand, researchers often do not have the resources for obtaining a sufficient number of ratings that are worth to be publicly released.

Because of the shortage of public datasets, practitioners have started to rely on synthetic ratings in order to conduct their offline experiments~\cite{Yu2012}. An obvious advantage of such an approach is that it enables the creation of rating datasets with an arbitrary number of users and items at a limited cost of dataset acquisition. However, the results obtained from such experiments may be questionable, as the generated datasets are usually not capable of capturing the characteristics of a particular domain of interest~\cite{Montaner2004}. For example, different generative approaches only rely on descriptive statistics, like mean and standard deviation, and, for this reason, they fail to mimic the individual behavior of a user.

In this work, we propose a novel approach for automatically generating synthetic datasets with a configurable number of users leveraging on a reference dataset that is used as the seed of the process and that encodes the peculiarities of a domain of interest. Such a generative method can be exploited to create different rating datasets containing users that exhibit behaviors similar to the ones available in the reference dataset. However, the synthetic users do not have a direct relation with the real users and, therefore, no private or commercially sensible information is leaked. At the same time, because the number of synthetic users is configurable, the generated dataset can be exploited to conduct scalability tests in a realistic way and to train recommendation algorithms using reinforcement learning approaches.

More formally, we aim to provide an answer to the following research questions.

\begin{description}
\item[RQ1] What is the impact of using a synthetic dataset instead of a real one on the results of an offline experiment in the context of recommender systems?
\item[RQ2] Can a generative approach be exploited to create a synthetic dataset that exhibits properties similar enough to the ones of a real dataset?
\item[RQ3] To what extend this method can be consistently applied to datasets from different domains and of different sizes?
\end{description}

The remainder of this paper is structured as follows: in Section~\ref{sec:related} we review related works and we compare them to our proposal. In Section~\ref{sec:generation} we introduce the generative approach for creating synthetic datasets, while in Section~\ref{sec:experiments} we describe the experimental setup designed to validate it. We present and discuss the results in Section~\ref{sec:results} and, in Section~\ref{sec:conclusion}, we provide the conclusions.

\section{Related work}
\label{sec:related}

Synthetic datasets are commonly used in literature to assess the performance of database systems or to study the behavior of data mining algorithms. For example, Agrawal et al.~\cite{Agrawal1994} created a generator of retail transactions intended for the evaluation of association rule algorithms, while Houkj{\ae}r et al.~\cite{Houkjaer2006} introduced a software capable of creating relational data for benchmarking purposes. Such tools can generate realistic data in terms of their statistical distributions, which can be empirically learned for existing datasets or provided by a researcher using specialized languages.

Similar approaches have been also explored in the field of recommender systems, usually because of the lack of public datasets with the required characteristics. Tso et al.~\cite{Tso2006} created a synthetic data generator for evaluating context-aware recommenders based on Dirichlet and Chi-square distributions. The metric of information entropy is then exploited to control the randomness of the synthetic data. A similar method has been discussed by Pasinato et al.~\cite{Pasinato2013}: their intuition is to represent the heterogeneous rating behaviors of the users with different statistical distributions.

Manouselis et al.~\cite{Manouselis2008} presented a tool, named CollaFis, capable of creating synthetic ratings for the evaluation of either single-criteria or multi-criteria recommender systems. The users of CollaFis need to specify the characteristics of the generated data, like the number of users, items, and criteria. A common aspect of all the previously mentioned methods is that researchers are required to choose and configure the statistical distributions that are exploited to generate the artificial datasets. However, the main problem of such an approach is that it is impossible to predict the real behavior of many different users with a few statistical distributions~\cite{Montaner2004}.

Another possible line of research is related to the imitation of a real collection of preferences. For example, CarmenRodr{\'i}guez Hern{\'a}ndez et al.~\cite{CarmenRodriguez-Hernandez2017} developed a software, DataGenCARS, for creating artificial ratings using a set of parameters provided by the user or inferred from a reference dataset. However, we argue that statistics computed at a global level are not informative enough to create a synthetic dataset, as they are not able to capture the different behaviors of the various groups of users.

\section{Dataset generation}
\label{sec:generation}

Our approach for generating synthetic datasets starting from a reference dataset consists of two steps. In the first one, it is necessary to analyze an existing collection of user preferences in order to obtain an accurate representation of the domain of interest. Then, in the second one, it is possible to exploit such a representation for creating different generated datasets.

We argue that only relying on a few statistical distributions computed empirically at a global level from an existing dataset or specified by a researcher is not sufficient to realistically simulate the individual tastes of human beings~\cite{Montaner2004}. Such methods would lead to the creation of datasets with users having no individual preferences, thus making the task of any recommender system nearly impossible.

For this reason, we included a preliminary clustering phase as part of the first step in order to group the users in a fixed number of communities. The individual rating behaviors, represented by different statistical distributions, are learned for each community and then exploited during the sampling phase.

For simplicity, we assume that each user can only express positive preferences about the items available in the system. However, this approach can also be exploited to simulate datasets with ratings expressed on a more complex scale by repeating these steps for each rating value and then by merging the results.

In the following, we detail the user clustering and distribution learning process and the rating sampling algorithm.

\subsection{User clustering and distribution learning}

We represent each user $\upsilon \in \mathcal{U}$ from the reference dataset as a vector with length equal to the number of items $|\mathcal{I}|$. The component $\hat{\upsilon}_i$ of such a vector is equal to $1$ if the user $\upsilon$ expressed a positive rating $\rho$ about the \textit{i}-th item of the catalog, otherwise it is equal to $0$.

Given this data structure, we decided to apply the K-means clustering algorithm~\cite{Hartigan1979} to group together users who liked a similar set of items in $K$ different clusters. The value of $K$ needs to be empirically selected by the experimenter because, in general, it depends on the characteristics of the reference dataset.

Every cluster identifies a different community of users. For generating a dataset similar to the reference one, it is necessary to know how many users belong to each community and what are the item preferences associated with them. More in detail, we create the following empirical distributions from the reference ratings:

\begin{itemize}
    \item $P^C$, how users are distributed in $K$ clusters;
    \item $P^U_k$, how ratings are distributed in $|\mathcal{U}|$ users for each cluster;
    \item $P^I_k$, how ratings are distributed in $|\mathcal{I}|$ items for each cluster.
\end{itemize}

Note that only the first distribution is global, while the second and the third ones are associated with a cluster.

The distribution $P^C$ represents the probability of assigning a user to a certain cluster and it is computed by counting the number of users per cluster. The distribution $P^U_k$ represents the probability of finding a certain number of ratings per user in the cluster $k$ and it is computed by counting the number of ratings per user. Finally, the distribution $P^I_k$ represents the probability of finding a certain number of ratings per item in the cluster $k$ and it is computed by counting the number of ratings per item.

The user clustering and distribution learning process is formalized in Algorithm~\ref{alg:clustering}. Its output is represented by the previously mentioned empirical distributions.

\begin{algorithm}
\caption{User clustering and distribution learning, given a reference dataset and the number of clusters.}
\label{alg:clustering}
\begin{algorithmic}[1]
\REQUIRE $\mathcal{U} \neq \{\emptyset\} \land K > 0 \land K \leq |\mathcal{U}|$
\STATE $\mathcal{C} \gets \textrm{K-means}(\mathcal{U}, K)$
\STATE $P^C \gets P(\upsilon \in \mathcal{C}_k)$
\FORALL{$k \in \{1, \dotsc, K\}$}
\STATE $P^U_k \gets P(\rho_{\upsilon} | \upsilon \in \mathcal{C}_k)$
\STATE $P^I_k \gets P(\rho_{\iota} | \iota \in \mathcal{I}_{\upsilon} \land \upsilon \in \mathcal{C}_k)$
\ENDFOR
\RETURN $P^C,\ P^U_k,\ P^I_k$
\end{algorithmic}
\end{algorithm}

\subsection{Rating sampling}

Starting from the empirical distributions obtained from Algorithm~\ref{alg:clustering}, it is possible to generate a synthetic dataset by applying to them a sampling function $\sigma$. In the following, we assume that $\sigma$ is the weighted random sampling function.

As discussed in Section~\ref{sec:introduction}, the experimenter can select the number of users available in the generated dataset. This value, called $U$, is an input of the rating sampling algorithm, together with the probability distributions. The synthetic dataset can also have the same number of users in the reference dataset, that is $U = |\mathcal{U}|$.

Firstly, each generated user $u$ is assigned to a cluster $k$ from the reference dataset, according to the distribution of users per cluster. Then, the number of ratings $I$ for that user is selected considering the distribution of ratings per user in the cluster $k$. Finally, $I$ items are sampled without replacement ($\hat{\sigma}$) from the distribution of ratings per item in the cluster $k$. Thus, the number of user ratings and her liked items are associated with a particular community of users.

The rating sampling procedure is formalized in Algorithm~\ref{alg:generate}.

\begin{algorithm}
\caption{Rating sampling, given the required number of users and the distributions computed in Algorithm~\ref{alg:clustering}.}
\label{alg:generate}
\begin{algorithmic}[1]
\REQUIRE $U > 0, P^C, P^U_k, P^I_k$
\STATE $\mathcal{R} \gets \{\emptyset\}$
\FORALL{$u \in \{1, \dotsc, U\}$}
\STATE $k \gets \sigma(P^C)$
\STATE $I \gets \sigma(P^U_k)$
\FORALL{$i \in \{1, \dotsc, I\}$}
\STATE $\rho_{u, i} \gets \hat{\sigma}(P^I_k)$
\STATE $\mathcal{R} \gets \mathcal{R} \cup \{\rho_{u, i}\}$
\ENDFOR
\ENDFOR
\RETURN $\mathcal{R}$
\end{algorithmic}
\end{algorithm}

\section{Experimental setup}
\label{sec:experiments}

We compared the results obtained from the evaluation of different recommenders conducted on popular datasets typically exploited in literature with the ones computed in the same experimental conditions using various collections of synthetic preferences generated starting from them using multiple techniques.

In fact, we claim that a synthetic dataset can be successfully used during an evaluation campaign if the behavior of the recommender systems under analysis is similar to one that it would be possible to observe with the reference dataset. Thus, almost all the possible pairs of recommenders should exhibit the same relation of order for a given dimension and lead to similar conclusions.

Furthermore, we investigated what is the impact of the parameter $K$ on the results of the evaluation, in order to understand how to empirically select the most appropriate value for it.

In our experiments, we utilized Random, Most Popular, User KNN, BPRMF, and WRMF recommendation algorithms and the metrics of precision, recall, and NDCG as defined in the evaluation framework RecLab~\cite{Monti2018}. Regarding the user preferences, we exploited the binarized versions of the MovieLens~100K, MovieLens~1M, and LastFM~\cite{Cantador2011} datasets. We considered as positive all ratings with a value higher than $3$ for MovieLens and than $0$ for LastFM. We relied on the default values of the evaluation framework for all other experimental parameters: we followed a random splitting protocol with a test set size equal to the $20\%$ of all available ratings and we recommended $10$ items for each test user.

From the aforementioned reference datasets we generated their synthetic versions exploiting the procedure described in Section~\ref{sec:generation}. We considered $U$ equal to the number of users originally available, in order to compare datasets of similar size. Furthermore, we also created three baseline synthetic collections with the same number of ratings by not applying the user clustering phase. All the users of such baselines exhibit the same rating behavior, similarly to the approach described in \cite{CarmenRodriguez-Hernandez2017}. In Table~\ref{tab:stats}, we report different statistics regarding the baseline, generated, and reference datasets.

\begin{table}
\caption{The total number of users, items, and ratings available in the datasets under consideration.}
\label{tab:stats}
\begin{tabular}{@{}llrrr@{}}
\toprule
Dataset & Version & Users & Items & Ratings \\ \midrule
MovieLens 100K & Baseline & 942 & 1,374 & 55,375 \\
MovieLens 100K & Generated & 942 & 1,332 & 53,915 \\
MovieLens 100K & Reference & 942 & 1,447 & 55,375 \\ \midrule
MovieLens 1M & Baseline & 6,038 & 3,463 & 575,281 \\
MovieLens 1M & Generated & 6,038 & 3,457 & 584,101 \\
MovieLens 1M & Reference & 6,038 & 3,533 & 575,281 \\ \midrule
LastFM & Baseline & 1,888 & 13,342 & 92,834 \\
LastFM & Generated & 1,892 & 13,442 & 92,510 \\
LastFM & Reference & 1,892 & 17,632 & 92,834 \\ \bottomrule
\end{tabular}
\end{table}

\section{Results}
\label{sec:results}

In this section, we first discuss the impact of the number of user communities on the evaluation results, then we present a comparison between exploiting the synthetic and the reference datasets.

\subsection{Number of user communities}
\label{sec:communities}

\begin{table}
\caption{The values of precision obtained with the generated versions of the MovieLens~100K dataset by varying $K$.}
\label{tab:clusters}
\begin{tabular}{@{}lrrrr@{}}
\toprule
Dataset & Most Popular & User KNN & BPRMF & WRMF \\ \midrule
K = 5 & 0.088449 & 0.099890 & 0.078768 & 0.091749 \\
K = 10 & 0.095793 & 0.124595 & 0.102805 & 0.111974 \\
K = 50 & 0.098378 & 0.133946 & 0.103243 & 0.133838 \\
K = 100 & 0.102415 & 0.150494 & 0.115587 & 0.149945 \\
K = 200 & 0.099672 & 0.154158 & 0.122538 & 0.164114 \\ \bottomrule
\end{tabular}
\end{table}

For studying what is the impact of the value $K$ on the results of an evaluation conducted with a synthetic dataset, we computed the measure of precision on different synthetic versions of the MovieLens~100K dataset created with $K = \{5,\ 10,\ 50,\ 100,\ 200\}$. We report the numerical outcomes of this experiment in Table~\ref{tab:clusters}.

We also observed that it is possible to obtain similar results by considering other datasets and metrics. As expected, the values of precision for all the algorithms but the Random and Most Popular approaches improve by increasing the number of available clusters. However, this relationship is not linear, as doubling its value from $100$ to $200$ only slightly improves the results.

We empirically observed that reasonable values for $K$ could be $100$ or $200$. In Section~\ref{sec:datasets}, we will assume that $K = 200$.

Therefore, we can provide an answer to \textbf{RQ1} by observing that the impact of using a synthetic dataset in an evaluation campaign can be mitigated if we are able to simulate a sufficient number of heterogeneous user communities.

\begin{table*}
\caption{The results obtained with the baseline, generated, and reference versions of MovieLens~100K.}
\label{tab:movielens-100k}
\begin{tabular}{@{}llll|lll|lll@{}}
\toprule
 & \multicolumn{3}{c}{Baseline dataset} & \multicolumn{3}{c}{Generated dataset} & \multicolumn{3}{c}{Reference dataset} \\ \midrule
Algorithm & Precision & Recall & NDCG & Precision & Recall & NDCG & Precision & Recall & NDCG \\ \midrule
Random & 0.009416 & 0.008877 & 0.009841 & 0.009847 & 0.008977 & 0.010022 & 0.007743 & 0.006300 & 0.008183 \\
Most Popular & 0.060065 & 0.053209 & 0.064384 & 0.099672 & 0.083875 & 0.110229 & 0.112759 & 0.102804 & 0.130632 \\
User KNN & 0.055952 & 0.050587 & 0.058744 & 0.154158 & 0.135917 & 0.169499 & 0.205234 & 0.221684 & 0.233362 \\
BPRMF & 0.045346 & 0.033628 & 0.048740 & 0.122538 & 0.106186 & 0.129742 & 0.182770 & 0.186838 & 0.198869 \\
WRMF & 0.047078 & 0.042876 & 0.048104 & 0.164114 & 0.144272 & 0.173916 & 0.221592 & 0.233235 & 0.250386 \\ \bottomrule
\end{tabular}
\end{table*}

\begin{table*}
\caption{The results obtained with the baseline, generated, and reference versions of MovieLens~1M.}
\label{tab:movielens-1m}
\begin{tabular}{@{}llll|lll|lll@{}}
\toprule
 & \multicolumn{3}{c}{Baseline dataset} & \multicolumn{3}{c}{Generated dataset} & \multicolumn{3}{c}{Reference dataset} \\ \midrule
Algorithm & Precision & Recall & NDCG & Precision & Recall & NDCG & Precision & Recall & NDCG \\ \midrule
Random & 0.004989 & 0.002733 & 0.004803 & 0.005712 & 0.002678 & 0.005580 & 0.005589 & 0.002862 & 0.005657 \\
Most Popular & 0.066483 & 0.037723 & 0.068991 & 0.101570 & 0.061565 & 0.107356 & 0.131982 & 0.082978 & 0.142782 \\
User KNN & 0.064708 & 0.035976 & 0.066573 & 0.129113 & 0.078716 & 0.136779 & 0.232082 & 0.172290 & 0.262018 \\
BPRMF & 0.057459 & 0.027145 & 0.059226 & 0.106948 & 0.057729 & 0.111058 & 0.199633 & 0.136378 & 0.218727 \\
WRMF & 0.053022 & 0.027763 & 0.055035 & 0.133723 & 0.080045 & 0.140670 & 0.227878 & 0.154425 & 0.252999 \\ \bottomrule
\end{tabular}
\end{table*}

\begin{table*}
\caption{The results obtained with the baseline, generated, and reference versions of LastFM.}
\label{tab:lastfm}
\begin{tabular}{@{}llll|lll|lll@{}}
\toprule
 & \multicolumn{3}{c}{Baseline dataset} & \multicolumn{3}{c}{Generated dataset} & \multicolumn{3}{c}{Reference dataset} \\ \midrule
Algorithm & Precision & Recall & NDCG & Precision & Recall & NDCG & Precision & Recall & NDCG \\ \midrule
Random & 0.000691 & 0.000656 & 0.000869 & 0.000532 & 0.000520 & 0.000548 & 0.000797 & 0.000825 & 0.000791 \\
Most Popular & 0.046281 & 0.046513 & 0.048636 & 0.052844 & 0.054597 & 0.056285 & 0.067906 & 0.068970 & 0.075406 \\
User KNN & 0.042614 & 0.043088 & 0.044308 & 0.101435 & 0.104549 & 0.113433 & 0.156057 & 0.160451 & 0.189487 \\
BPRMF & 0.039957 & 0.040543 & 0.041804 & 0.062307 & 0.064720 & 0.066835 & 0.075877 & 0.077336 & 0.087066 \\
WRMF & 0.032731 & 0.032974 & 0.033975 & 0.090324 & 0.092860 & 0.099598 & 0.160202 & 0.164468 & 0.193937 \\ \bottomrule
\end{tabular}
\end{table*}

\subsection{Synthetic and reference datasets}
\label{sec:datasets}

As anticipated in Section~\ref{sec:experiments}, we compared the evaluation results obtained when relying on the reference dataset and two synthetic datasets created with different approaches. We repeated this experiment with datasets of different sizes and from different domains in order to assess the generalizability of the results.

The results obtained with MovieLens~100K, MovieLens~1M, and LastFM are available in Table~\ref{tab:movielens-100k}, Table~\ref{tab:movielens-1m}, and Table~\ref{tab:lastfm} respectively.

We observe that in all experiments and for almost all the possible pairs of recommenders the relative order of the measures is the same between the generated and the reference datasets.

As expected, their values are lower when exploiting the synthetic ratings, as they do not represent real preferences. Nevertheless, they are still useful to identify the most promising recommendation techniques in a certain domain, while the results obtained with the baseline datasets cannot be exploited for such a purpose. 

With respect to \textbf{RQ2}, we can conclude that a generative approach capable of replicating the behaviors of different groups of users can be used for creating realistic datasets. We also discovered, as an answer to \textbf{RQ3}, that our approach can be potentially applied to datasets from different domains and of different sizes.

\section{Conclusion and future work}
\label{sec:conclusion}

In this paper, we have discussed a method for generating synthetic datasets with an arbitrary number of users starting from existing collections of preferences. Differently from the approaches already available in literature, we propose to first model user communities in order to generate more realistic ratings that can be successfully exploited during an evaluation campaign.

We empirically verified that the outcome of an offline comparison among different recommender systems conducted exploiting the generated datasets is consistent with the results obtained when using the reference datasets, provided that a sufficient number of user clusters is selected. This finding could encourage private companies to publicly release synthetic datasets created from internally available data without the fear of violating the privacy of their users or of exposing commercially sensible information.

As future work, we would like to explore additional methods for creating synthetic datasets. We believe that Generative Adversarial Networks (GANs) could be successfully exploited for this task, as they are already used to generate fake images starting from real ones~\cite{Goodfellow2014}. Such approaches would require the definition of a way for representing the preferences of a user similarly to an image.

\clearpage\balance
\bibliographystyle{ACM-Reference-Format}
\bibliography{references} 

\end{document}